\documentclass[preprint, sort&compress, number]{elsarticle}

\usepackage[T1]{fontenc}
\usepackage{graphicx}
\usepackage{amsmath}

\begin{document}

\begin{frontmatter}

\title{Structural studies of $^1$H-containing liquids by polarized neutrons: chemical environment and wavelength dependence of the incoherent background}


\author[1]{L\'aszl\'o Temleitner}
\author[1,2]{L\'aszl\'o Pusztai \corref{cor1}}
\ead{pusztai.laszlo@wigner.hu}
\author[3]{Gabriel J. Cuello}
\author[3]{Anne Stunault}

\cortext[cor1]{Corresponding author}

\affiliation[1]{organization={Institute for Solid State Physics and Optics, Wigner Research Centre for Physics}, addressline={Konkoly Thege {\'u}t 29-33.}, postcode={1121}, city={Budapest}, country={Hungary}}
\affiliation[2]{organization={International Research Organisation for Advanced Science and Technology (IROAST)}, addressline={Kumamoto University, 2-39-1 Kurokami, Chuo-ku}, postcode={860-8555}, city={Kumamoto City}, country={Japan}}
\affiliation[3]{organization={Institute Laue Langevin}, addressline={71 Avenue des Martyrs}, postcode={38000}, city={Grenoble}, country={France}}

\begin{abstract}

Following a demonstration of how neutron diffraction with polarization analysis may be applied for the accurate determination of the coherent static structure factor of disordered materials containing substantial amounts of proton nuclei (Temleitner et al., Phys. Rev. B 92, 014201, 2015), we now focus on the incoherent scattering. Incoherent contributions are responsible for the great difficulties while processing standard (non-polarized) neutron diffraction data from hydrogenous materials, hence the importance of the issue. Here we report incoherent scattering intensities for liquid acetone, cyclohexane, methanol and water, as  function of the $^1$H/H ratio. The incoherent intensities are determined directly by polarized neutron diffraction. This way, possible variations of the incoherent background due to the changing chemical environment may be monitored. In addition, for some of the water samples, incoherent intensities as a function of the wavelength of the incident neutron beam (at 0.4, 0.5 and 0.8 \AA ) have also been measured. It is found that in each case, the incoherent intensity can be described by a single Gaussian function, within statistical errors. The (full) width (at half maximum) of the Gaussians clearly depends on the applied wavelength. On the other hand, the different bonding environments of hydrogen atoms do not seem to affect the width of the Gaussian. 

\end{abstract}

\begin{graphicalabstract}
\resizebox{0.96\textwidth}{!}{\includegraphics{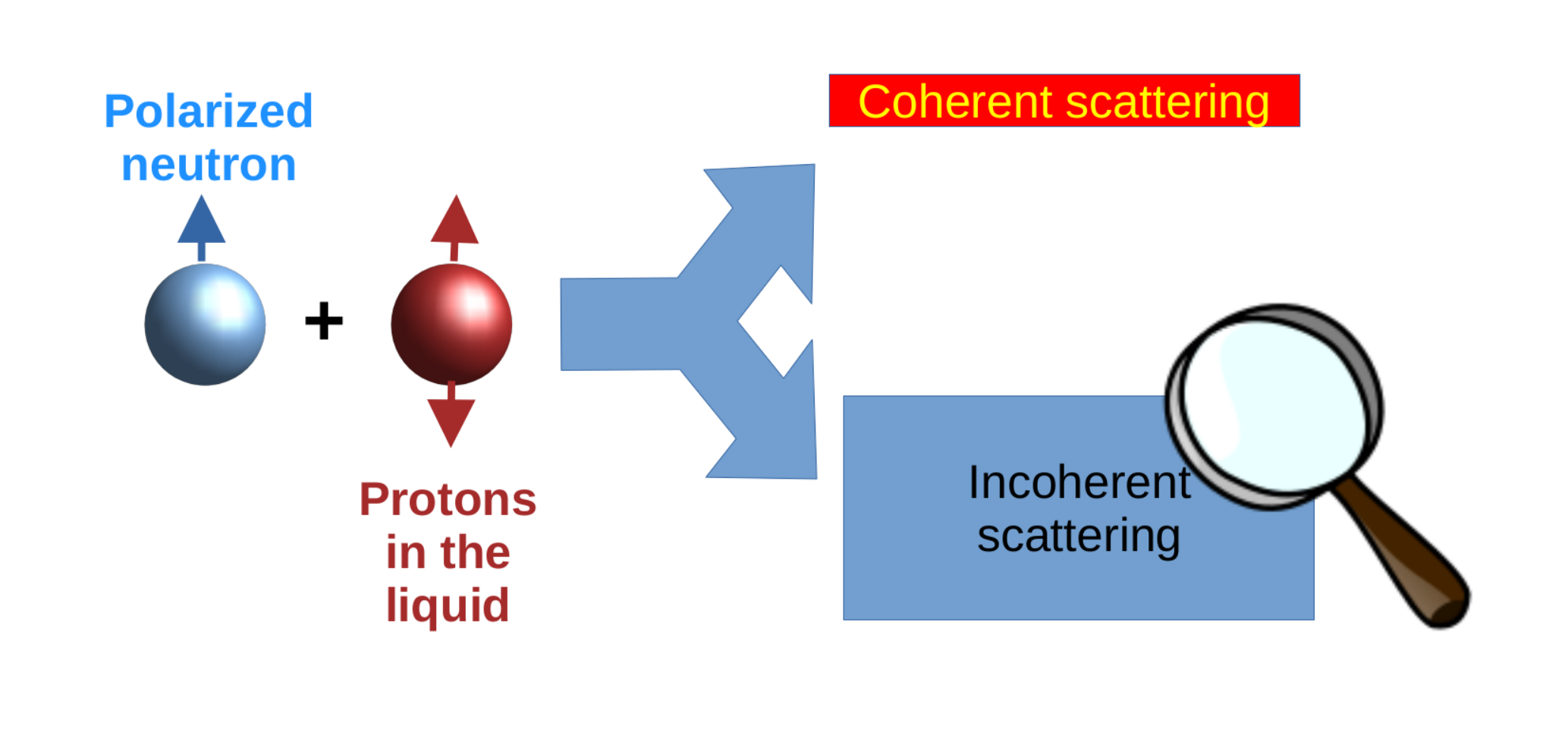}}
\end{graphicalabstract}

\begin{highlights}
    \item Incoherent neutron scattering contributions have been measured for hydrogen containing liquids.
    \item Incoherent scattering contributions can be fitted well by a Gaussian curve and a constant.
    \item The FWHMs of the Gaussians depend on the wavelength, but not on the hydrophobic/phylic character of the proton.
\end{highlights}





\end{frontmatter}

\section{Introduction}

The primary reason why the determination of the structure of hydrogen-containing materials is still a burdensome mission is, quite simply, the presence of \textit{hydrogen}. X-ray diffraction is hardly sensitive to hydrogen and can provide information only on oxygen-related pair correlations (see, e.g., Ref\cite{skinner}). In neutron diffraction, separating the three partial contributions would be possible via the contrast variation between $^1$H and $^2$H, via the so-called 'H/D isotopic substitution' \cite{thiessen_82,soper_97}. This, in principle, allows us to derive the most detailed information on the microscopic structure of hydrogenous (i.e., containing $^1$H) systems, due to the large difference between their coherent neutron scattering lengths \cite{ncoherent}: $b_c^H=-3.7406fm$ versus $b_c^D=6.671fm$.

Reliable 'neutron weighted' total structure factors of hydrogenous samples with high  $^1$H content (ideally, containing only the hydrogen isotope $^1$H) would be decisive concerning, for instance, hydrogen-bonding (H-bonding) in water, alcohols, and aqueous mixtures. For this reason, numerous suggestions over the past 40 years have been made for the treatment of the huge 'incoherent background' (for informative figures, see, e.g., \cite{thiessen_82,temleitner_2015}), none of which has proven to be routinely applicable (for the various approaches, see \cite{powles_72,blum_76,powles_79,granada_87a,dawidowski_94,palomino_07}).

Genuine improvement could only be expected from accurate experimental determination of the incoherent contributions from $^1$H, over a wide enough momentum transfer range; this, unfortunately, has proven to be impossible until recently \cite{temleitner_2015,palomino_2015}. 

Spin-incoherence can, in principle, be tackled by separating the coherent and incoherent parts of the measured diffraction signal; this can be realized by using polarized neutrons (see, e.g., Ref.~\cite{polar_textbook}). By measuring total scattering intensities as recorded by "non-spin-flip" (NSF) and by "spin-flip" (SF) modes of the instrument, the coherent and spin-incoherent intensities can be extracted using the following formulae:

\begin{equation}\label{eq:coh}
I_{coh}(Q)=I^{NSF}(Q) - \frac{1}{2} I^{SF}(Q)
\end{equation}

and

\begin{equation}\label{eq:incoh}
I_{incoh}(Q) = \frac{3}{2} I^{SF}(Q).
\end{equation}

Still, potentialities of polarized neutron diffraction have only been little exploited in this field. A possible reason for this is that available instruments provide data over only narrow momentum transfer ranges (see, e.g. \cite{temleitner_07}), so that traditional evaluation, involving direct Fourier-transformation, would not be applicable. 

In a recent publication\cite{temleitner_2015} the present authors have demonstrated that the determination of the coherent part of the scattering is a realistic possibility. Some further instrumental and data treatment advances have just appeared\cite{palomino_2015,stunault_2016}. In the present study, new (polarized neutron diffraction) measurements on liquid acetone, cyclohexane, methanol and water are presented, along with relevant experimental and data correction details. This time, the focus is put on the incoherent scattering contributions, as this kind of scattering is responsible for the huge background when standard neutron diffraction is applied on $^1$H-containing materials.

\section{Instrumental details and primary data treatment}

Diffraction experiments using polarized neutrons have been conducted on the D3 instrument \cite{D3-REFERENCE} installed on the hot source of the Institut Laue-Langevin (ILL; Grenoble, France). 


Liquid samples were put in a double-walled vanadium container (internal diameter: 8 mm, outer diameter: 10.7 mm), in order to minimize contributions from multiple scattering; the sample geometry was therefore that of a hollow cylinder. The experiments were performed at ambient pressure and temperature. Using the D3 instrument with 0.5~\AA{} wavelength neutrons\cite{cyclo_measured, water05_measured}, scattering intensities have been collected in both spin-flip and non-spin-flip modes over a uniquely wide momentum transfer range of 0.8-21~\AA$^{-1}$ (4-120 degrees in $2\Theta$). The hot neutron source of the Institut Laue-Langevin can provide a high flux of such short wavelength neutrons, which is the pre-requisite for studies like reported here. This outstanding coverage of the reciprocal space can be realized by making use of a Heussler-alloy polarizer and a $^3$He analyzer cell that contains spin-polarized nuclei \cite{surkau_97}. For studying wavelength dependence, two other monochromator setups have also been applied, so that neutron beams with wavelengths of 0.4 and 0.8 {\AA} (corresponding to 1.4-27.4~\AA$^{-1}$ and 0.7-13.4~\AA$^{-1}$ ranges in terms of momentum transfer, respectively) have also been generated\cite{water0408_measured}. We emphasize that, although it has not been designed for liquid diffraction, the D3 instrument at present is the only one in the world where studies of this kind may be conducted. 

Samples with the highest $^1$H content have been investigated for somewhat longer time than those dominantly with $^2$H, so that statistics of the coherent signals would be comparable. Still, the measuring time of about 24 hours for fully protonated samples provided statistics somewhat poorer than hoped for. The usual corrections\cite{heil_02} for the time-dependent polarization efficiency for the cells have also been carried out before further data processing. Coherent and spin-incoherent contributions to the total scattering have been separated in the usual manner\cite{polar_textbook} by eqs. \ref{eq:coh} and \ref{eq:incoh}.

\section{Results and discussion}

\begin{figure}[ht]
\begin{center}
\rotatebox{0}{\resizebox{1.0
\textwidth}{!}{\includegraphics{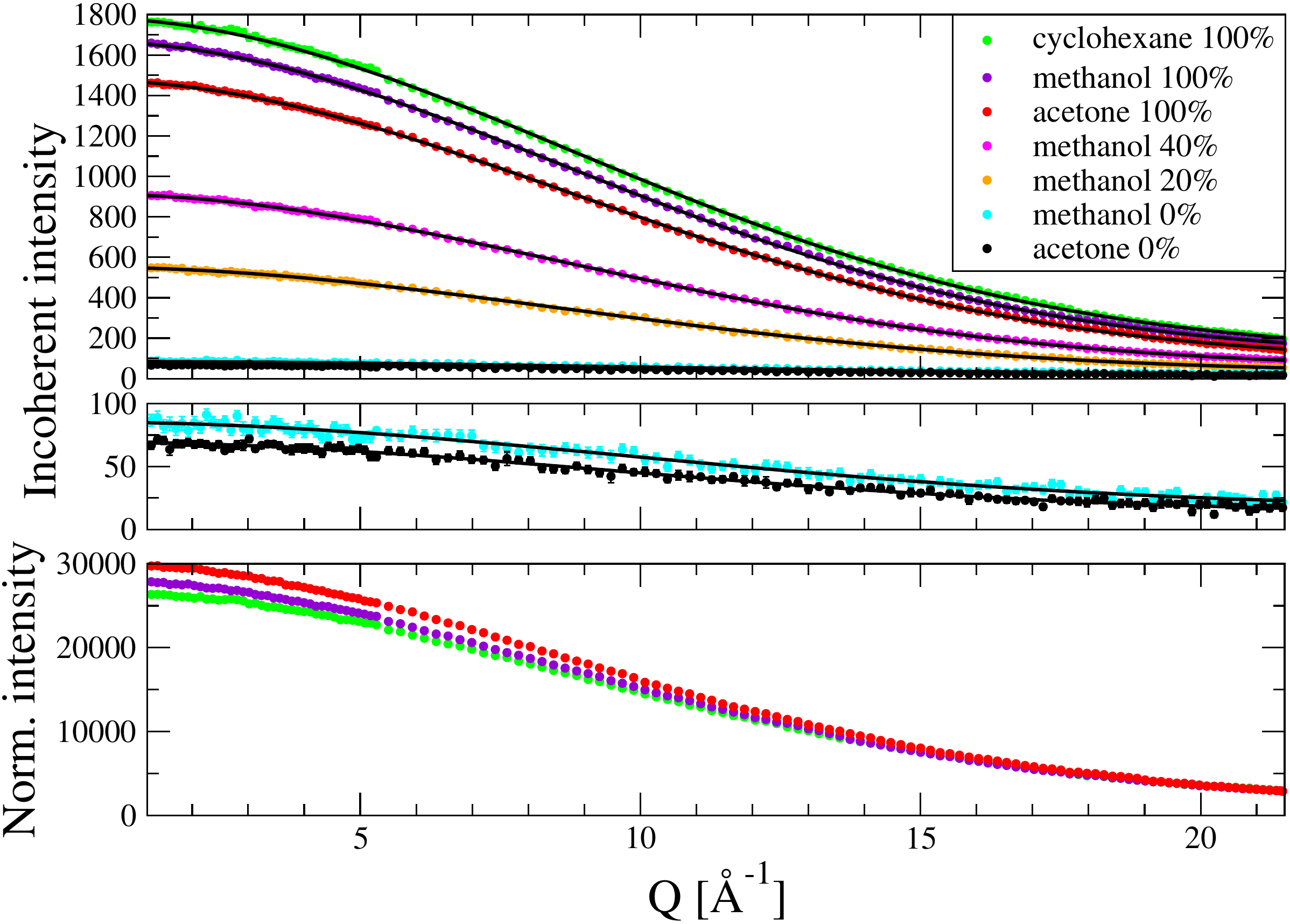}}}
\caption{\label{fig:incoh-1}
Upper and middle panels: Incoherent intensities, obtained by multiplying the measured 'spin-flip' intensities by 3/2 (see eq. \ref{eq:incoh}), for (from top to bottom) cyclohexane (fully protonated), methanol (fully protonated), acetone (fully protonated), 40, 20 and 0 \% protonated methanol (in this order), and finally, fully deuterated acetone. Individual Gaussian fits are also shown; they are hardly distinguishable from the measured curves. Lower panel: Normalized incoherent intensities for 1 {\AA$^{-3}$} hydrogen density for fully protonated methanol, acetone and cyclohexane. The aim of normalizing intensities was to find a 'general' curve that would depend only on the density of protons. Although differences between normalized curves appear to be small, they are significant (see text).
}
\end{center}
\end{figure}

We start with the presenting experimental data taken for the investigation of the (possible) dependence on the chemical environment and hydrogen number density ($\rho_H$, Table \ref{tab:hnumberdensity}).
\begin{table}[ht]
\begin{center}
\footnotesize
\begin{tabular}{||c|cccc||}
\hline \hline
Compound & cyclohexane & water & methanol & acetone \\
\hline
$\rho_H$ [\AA$^{-3}$] & 0.06689 & 0.06665 & 0.05954 & 0.04914 \\
\hline \hline
\end{tabular}
\caption{\label{tab:hnumberdensity} 
Hydrogen number densities ($\rho_H$) of the studied materials}
\end{center}
\end{table}
In the liquids chosen for this purpose, cyclohexane contains only C and H atoms; acetone has H-atoms bound only to carbon (in methyl groups), but an oxygen atom is also present in its molecules. Molecules of methanol have H-atoms both in hydroxyl- and methyl groups, whereas water molecules consist of two 'fused' hydroxyl groups. Figure \ref{fig:incoh-1} shows the incoherent intensities, which are directly proportional to the spin incoherent cross sections, up to 21~\AA$^{-1}$ (neutron wavelength: 0.5 {\AA}).  

\begin{table}[ht]
\begin{center}
\footnotesize
\begin{tabular}{||cc|cccc||}
\hline \hline
Compound & $^1$H/H [\% ]& Constant term & Intensity & FWHM [\AA$^{-1}$] & R$_{wp}$ [\%] \\
\hline
cyclohexane & 100 & 1807(22) & 545205(1052) & 20.59(3) & 0.6 \\
\hline
methanol & 100 & 1804(19) & 566837(884) & 20.29(2) & 0.5 \\
 & 40 & 862(14) & 314849(689) & 20.40(3) & 0.7 \\
 & 20 & 500(13) & 189798(666) & 20.36(6) & 1.3 \\
 & 0 & 279(21) & 28379(1068) & 23.12(55) & 5.3 \\
\hline
acetone & 100 & 1695(21) & 614805(987) & 20.36(3) & 0.6 \\
 & 0 & 260(19) & 27184(902) & 22.12(45) & 4.62 \\
\hline \hline
\end{tabular}
\caption{\label{tab:diffh_parms} 
Adjusted parameters {\it Intensity}, {\it FWHM} and {\it Constant term}, together with goodness-of-fit (R$_{wp}$) values while fitting the measured incoherent intensities divided by the hydrogen number density ($\rho_H$) of cyclohexane, methanol and acetone samples by 0.5 \AA{} neutrons. }
\end{center}
\end{table}

At first sight, incoherent intensities seem to be proportional to the density of protons in the samples. As it can be seen in the lower panel of Figure \ref{fig:incoh-1} and numerically in Table \ref{tab:diffh_parms}, this presumption is almost valid: when the three incoherent intensities of fully protonated compounds are normalized to the proton content then only small differences remain between the signals. Note, however, that if we wished to use such a 'general' shape for background corrections to intensities measured without polarization analysis then even such small differences would prove to be crucial (cf. Figure 4 of Ref. \cite{temleitner_2015}). Therefore the hope for being able to provide a 'generalized' curve for the difficult incoherent inelastic background corrections had to be abandoned for the time being. Nevertheless, the hydrogen number density ($\rho_H$) of each material have been taken into account as a fixed multiplication factor in the functional form.

During our initial investigations on liquid water \cite{temleitner_2015}, it has become clear that a single Gaussian and an additional constant were perfect for fitting the measured signals for all the various mixtures of light and heavy water, within errors. Here, this statement is proven to be valid for a wider class of materials (see Figure \ref{fig:incoh-1}).
 
The exact form is: 
\begin{equation}\label{eq:fitfunc}
 I_{incoh}(Q) = \rho_H \left( Intensity\frac{2}{FWHM}\sqrt{\frac{\ln (2)}{\pi}}*\exp\left\{-4\ln (2) \left( \frac{Q}{FWHM}\right)^2\right\} + Constant \right)
\end{equation}

The above expression operates with only 3 adjustable parameters: {\it intensity}, {\it FWHM} and a {\it constant}. Note that this functional form, which was introduced in Ref. \cite{temleitner_2015} earlier, has not been suggested explicitly by any of the corresponding studies of the incoherent background\cite{powles_72,blum_76,powles_79,granada_87a,dawidowski_94,palomino_07}. On the other hand, the usefulness of Gaussian-like background corrections was mentioned in an extensive review\cite{fischer_06}; moreover, the incoherent inelastic background has routinely been fitted for $^1$H containing materials (see, e.g., Ref.\cite{gabriel-1}) while using the D4 diffractometer\cite{D4} at the Institut Laue-Langevin.

Out of the three fitting parameters, here the FWHM will be scrutinized (Table \ref{tab:diffh_parms}). For the proton-containing samples, its value fluctuated between 20.3 and 20.6 {\AA}$^{-1}$, with no any noticeable trend. These values are in full agreement with what was found for mixtures of light and heavy water (see Ref. \cite{temleitner_2015}). For the two fully deuterated samples, the FWHM-s were ca. 22 and 23 {\AA}$^{-1}$ (for pure heavy water, a value just above 21 {\AA}$^{-1}$ was found \cite{temleitner_2015}); note that, due to the nearly flat curves and insufficient statistics, FWHM-s for fully deuterated samples must be considered as much less reliable. 

It is striking that for all samples that contain any amount of protons, there seems to be a 'generic' FWHM parameter, whereas for the fully deuterated ones, there is a different 'generic' FWHM. That is, the widths of Gaussians are the same for a given isotope within the assumed uncertainties. With this notion, it should be noted that the FWHM values reported for methanol samples where two isotopes of hydrogen are present ($^1$H and $^2$H), the exact shape of the incoherent intensity curves should be taken as combinations of two Gaussians, having slightly different FWHM-s. The reason why this seems to be unimportant during the fitting procedure is that the amount of incoherent scattering from protons overwhelm that of from deuterons, so in practice, the deuterium content has to be taken into account only if it is close to 100 \% in a sample.

As noted above, the actual incoherent intensities in Figure \ref{fig:incoh-1} seemed, at first sight, to simply depend on the number density of protons in the liquids, which suggestion could not have been realized well enough. At the moment, insufficient corrections for multiple scattering effects are thought to be the main reason for the differences observed: properly elaborated correction procedures should be able to take into account the (expectedly, strong) dependence on the density of protons. For polarized beams, there is an additional problem: for instance, two consecutive 'spin-flip' events would appear as a single 'non-spin-flip' event, thus deceiving data acquisition. Development of an improved correction scheme is underway, along the path outlined in Ref. \cite{palomino_2015}.

\begin{table}[ht]
\begin{center}
\footnotesize
\begin{tabular}{||c|cccc||}
\hline \hline
$^1$H$_2$O/H$_2$O [\% ]& Constant term & Intensity & FWHM [\AA$^{-1}$] & R$_{wp}$ [\%] \\
\hline
100 & 3527(47) & 329716(1481) &  14.89(4) & 0.5 \\
92 & 3485(150) &  292743(1555) &  14.48(5) & 0.7 \\
70 & 2778(42) &  244589(1302) &  14.41(5) & 0.8 \\
55 & 2271(38) &  221248(1217) &  14.45(5) & 0.9 \\
38 & 1783(36) &  171698(1145) &  14.28(6) & 1.1 \\
\hline \hline
\end{tabular}
\caption{\label{tab:08_parms} 
Adjusted parameters {\it Intensity}, {\it FWHM} and {\it Constant term} (see eq. \ref{eq:fitfunc}), together with goodness-of-fit (R$_{wp}$) values while fitting the measured incoherent intensities
of water samples by 0.8 \AA{} neutrons.}
\end{center}
\end{table}

\begin{figure}[ht]
\begin{center}
\rotatebox{0}{\resizebox{1.0
\textwidth}{!}{\includegraphics{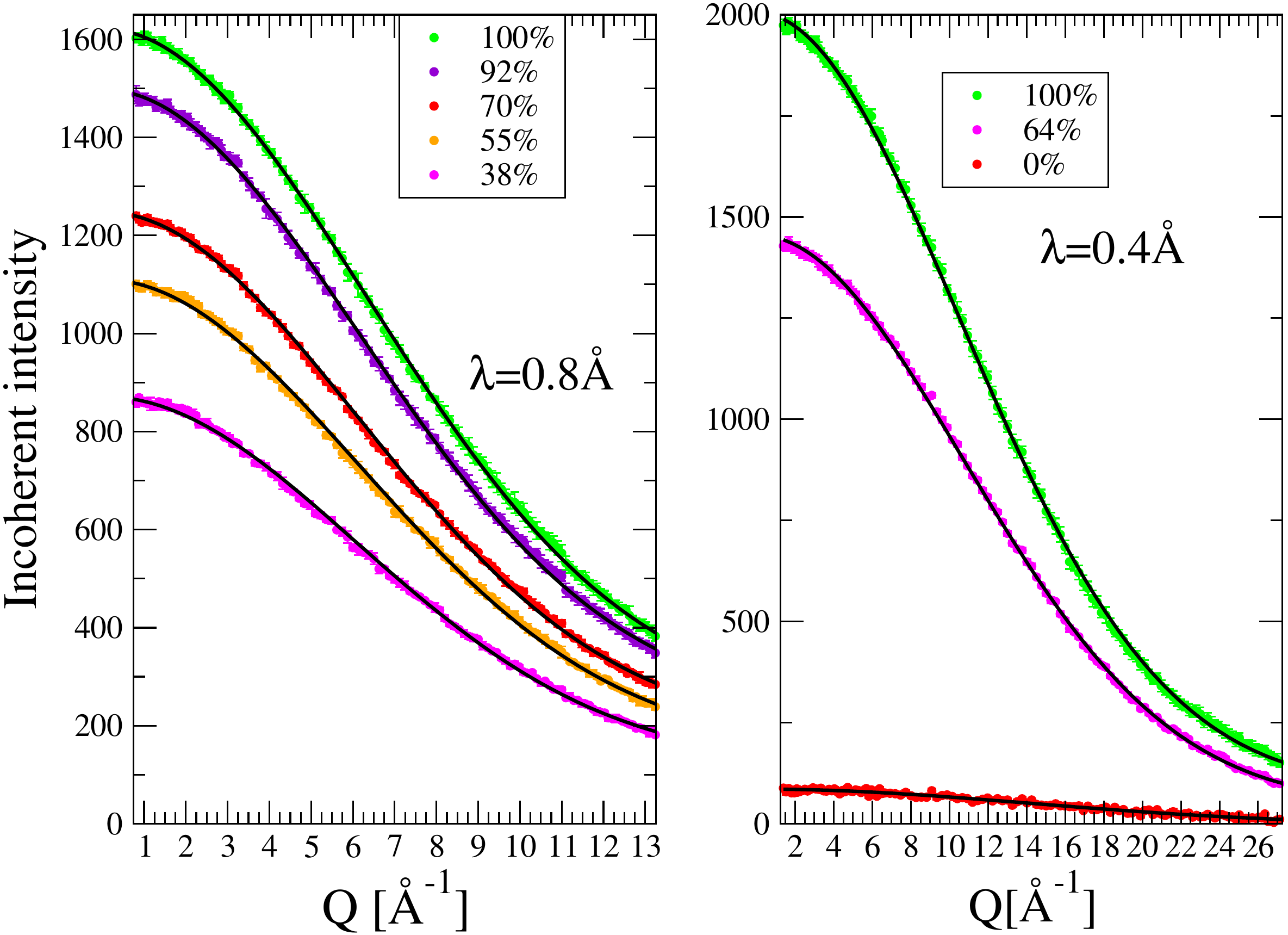}}}
\caption{\label{fig:incoh-2}
(a) Incoherent intensities for mixtures of light and heavy water at a neutron wavelength of 0.8 {\AA}, obtained by multiplying the measured 'spin-flip' intensities by 3/2 (see eq. \ref{eq:incoh}), for  light water contents of (from top to bottom) 100, 92, 70, 55 and 38 \%. Individual Gaussian fits are also shown. (b) Incoherent intensities for mixtures of light and heavy water at a neutron wavelength of 0.4 {\AA}, for  light water contents of (from top to bottom) 100, 64 and 0 \% (for the extremely long measuring times, we had no opportunity to take more data at this wavelength). Individual Gaussian fits are also shown.
}
\end{center}
\end{figure}

\begin{table}[ht]
\begin{center}
\footnotesize
\begin{tabular}{||c|cccc||}
\hline \hline
$^1$H$_2$O/H$_2$O [\% ]& Constant term & Intensity & FWHM [\AA$^{-1}$] & R$_{wp}$ [\%] \\
\hline
100 & 1284(41) & 759910(2500) & 24.78(6) & 0.7 \\
64 & 655(26) & 570111(1526) & 25.29(5) & 0.9 \\
0 & -45(50)  & 47245(3362) & 33.29(131) & 8.5 \\
\hline \hline
\end{tabular}
\caption{\label{tab:04_parms} 
Adjusted parameters {\it Intensity}, {\it FWHM} and {\it Constant term} (eq. \ref{eq:fitfunc}), together with goodness-of-fit (R$_{wp}$) values while fitting the measured incoherent intensities
of water samples by 0.4 \AA{} neutrons.}
\end{center}
\end{table}
Since it has not been clear previously whether the parameters of the Gaussian functions (i.e., the exact shape of the incoherent contribution) depend on the wavelength of the neutron beam \cite{temleitner_2015}, we have made use of three different setups of the D3 instrument and generated polarized neutron beams with wavelengths of 0.4, 0.5 and 0.8 \AA . While larger number of various light and heavy water mixtures have been measured at wavelengths of 0.8, only few for 0.4 {\AA} since the incoming intensity at this wavelength was much lower so counting took longer. Their measured incoherent intenisties are shown in Figure \ref{fig:incoh-2}. Also shown are the individual fits by Gaussians: the change in terms of the wavelength has no influence on the possibility of fitting incoherent intensities perfectly by Gaussian functions.

As far as the FWHM of the fitting Gaussians is concerned as a function of the wavelength, for $^1$H, they could be derived as ca. 14.5, 20 and 25 {\AA}$^{-1}$ for neutron wavelengths of 0.8, 0.5 (data for this wavelength are taken from Ref. \cite{temleitner_2015}) and 0.4 {\AA}, respectively. Concerning $^2$H, the FWHM-s are ca. 18.8, 21 and 33 {\AA}$^{-1}$ for wavelengths 0.8, 0.5 and 0.4 {\AA}, respectively.  That is, the wavelength dependence of the Gaussian widths is remarkably large, and the FWHM-s are roughly inversely proportional with the neutron wavelength. Also note that statistical uncertainties are much larger for fully deuterated samples, due to the much lower intensities of incoherent scattering. As a final remark here, it should be mentioned that the instrumental resolution also changes with the neutron wavelength, also inversely proportionally. However, since the variation of the Gaussian width with the wavelength is enormous, it is highly unlikely that resolution effects could account for wavelength dependence of the FWHM.

\section{Conclusions}

Exploiting potentialities of neutron diffraction with polarization analysis, we have been able to directly measure incoherent scattering contributions from a fairly wide range of liquids that contain hydrogen ($^1$H) with various chemical environments (in methyl-, methylene-, hydroxyl-groups, in water, methanol, acetone, and cyclohexane). For liquid water, we had the chance of varying the wavelength of the incoming neutron beam (0.4, 0.5 and 0.8 \AA ), as well. 

Having thus separated the incoherent contributions for a variety of protonated liquids, the precise removal of which would be essential for data analysis when using non-polarized neutron beams, the following statements can be made:
\begin{enumerate}
\item Each and every dataset may be described by a Gaussian, independently of the wavelength and the chemical environment of the H-atoms.
\item The FWHM of the Gaussians depends clearly on the wavelength (widest for 0.4 \AA , narrowest for 0.8 \AA ).
\item The FWHM of the Gaussians does not seem to depend on the chemical environment of the H-atoms (meaning that whether the H-atom is in a -CH$_3$ or in an -OH group, does not matter in this respect). This means that there is a 'universal' FWHM (but different for $^1$H and $^2$H) at each wavelength. 
\item The actual incoherent intensities at a given neutron wavelength are related primarily to the number of protons in the beam: this is a direct dependence on the number density of the protons in the liquid. The intensities also depend indirectly on the number density of the protons, since this is a major factor in determining multiple scattering contributions; efforts towards developing an accurate procedure for estimating the effects of multiple scattering are ongoing (see, e.g., \cite{palomino_2015,stunault_2016}).   
\end{enumerate}
The above findings are relevant from the point of view of structural studies of protonated materials, including aqueous mixtures, as well as organic compounds found in biochemistry and soft matter research. 

\section*{Acknowledgment}
LT and LP wish to thank the National Research, Development and Innovation Office of Hungary for financial support via grant No. KH130425. The contribution of LT is supported by the J\'anos Bolyai Research Scholarship of the Hungarian Academy of Sciences. Beamtime on the D3 instrument provided by the Institut Laue Langevin (Grenoble, France), under proposal numbers 6-02-536, 6-02-549 and 1-20-39, is gratefully acknowledged. We thank Ms. A. Szuja (Research Centre for Energy, Hungarian Academy of Sciences) for carefully preparing all mixtures of light and heavy water, and light and heavy methanol. 


\bibliography{polliquids-2019}

\end{document}